# Towards a Spreadsheet Engineering

V. R. Vemula, D. Ball, S. Thorne,
University of Wales Institute, Cardiff
V.R.Vemula@uwic.ac.uk  DBall@uwic.ac.uk  SThorne@uwic.ac.uk

**ABSTRACT**

*In this paper, we report some on-going focused research, but are further keen to set it in the context of a proposed bigger picture, as follows. There is a certain depressing pattern about the attitude of industry to spreadsheet error research and a certain pattern about conferences highlighting these issues. Is it not high time to move on from measuring spreadsheet errors to developing an armoury of disciplines and controls? In short, we propose the need to rigorously lay the foundations of a spreadsheet engineering discipline. Clearly, multiple research teams would be required to tackle such a big task. This suggests the need for both national and international collaborative research, since any given group can only address a small segment of the whole. There are already a small number of examples of such on-going international collaborative research. Having established the need for a directed research effort, the rest of the paper then attempts to act as an exemplar in demonstrating and applying this focus. With regard to one such of research, in a recent paper, Panko (2005) stated that: "…group development and testing appear to be promising areas to pursue." Of particular interest to us are some gaps in the published research record on techniques to reduce errors. We further report on the topics: techniques for cross-checking, time constraints effects, and some aspects of developer perception.*

**1. SPREADSHEET ENGINEERING**

Given the fact that spreadsheet modellers are not IS professionals, there has been significant effort to adapt existing software engineering principles to form a spreadsheet engineering discipline more sympathetic to spreadsheet modellers (Burnett *et al.* 2001, Burnett *et al.* 2003, Burnett *et al* 2004, Grossman 2002, Grossman and Ozluk 2004, Panko 2006, Nash and Goldberg 2005, Rajalingham *et al.* 2000). Some offer 'best practice' guidelines (Grossman 2002, Read and Batson, 1999, O'Beirne 2005) whilst others seek to develop a framework for spreadsheet engineering (Grossman and Ozluk 2004, Burnett *et al.* 2003, Burnett *et al* 2004, Rajalingham *et al.* 2000) or develop specific elements in a software lifecycle, such as testing (Panko 2006, Pryor 2004, Nash and Goldberg 2005, Yirsaw 2003)

Best practice guidelines in spreadsheets have proved difficult to settle on. Colver (2004) advocates that 'best practice' in spreadsheets is impossible to attain since adopting one approach to spreadsheet development often has negative side effects on the positive aspects of spreadsheet technology such as flexibility and speed of development. Other authors disagree, Read and Batson (1999) produced a detailed paper on spreadsheet best practice for organisations. This paper describes best practice from a systems development lifecycle approach, detailing best practices for planning, design, building, testing, maintenance and evaluation. This is compiled by the authors from years of experience gathered in Price Waterhouse Coopers (PWC). The actual best practice comes in the form of advice and guidelines for carrying out specific tasks as well as encouraging the reader to practice more general best practice, for example identifying stakeholders in the spreadsheet and conducting user acceptance testing respectively.

Grossman (2002) presents eight best practice principles based upon literature from spreadsheet modelling and a number of other related disciplines. Grossman highly recommends adopting best practice and presents evidence that doing so can significantly reduce error. O'Beirne (2005) draws from extensive experience to provide best practice in spreadsheets. The guidance offered comes in the form of both general, such as following a format when setting a spreadsheet up, and specific recommendations such as ensuring cell protection on cells with formulae.

**1.1 Framework for spreadsheet engineering**
Attempts have been made to modify and adapt frameworks in software engineering to substantiate spreadsheet engineering. This research is mainly concerned with identifying what is appropriate for spreadsheet modellers and is tested through experiments and field work.

Burnett *et al.* (2003) describes end user software engineering in the spreadsheet paradigm using assertions for debugging spreadsheets. It was discovered that the assertions helped the end users debug the spreadsheets, they caught more errors. Further, the participants routinely understood what the assertions meant and actually liked having them as a guide. This debugging was presented in the wider context of an iterative end user development life cycle.

Burnett *et al.* (2004) argues that since spreadsheet modellers are not IS professionals, it is more practical to employ a smaller feedback loop rather than provide a comprehensive traditional SDLC based methodology. The feedback loop incorporates the following: Interactive testing (testing while the user is modelling); Fault localisation (tool for locating faults after testing); Interactive assertions (monitoring values in the spreadsheet and alerting users to potential discrepancies) and motivational devices (Gets the user to participate in software engineering methods).

Grossman and Ozluk (2004) extend previous work on spreadsheet engineering principles, to give a more traditional adaptation of the SDLC and moves away from a best practice approach. This fresh approach gives consideration to the actual use of the final spreadsheet and recommends incorporating users into the development and holding a review of use after implementation.

**1.2 Evidence of Spreadsheet Errors**
Human errors are very common and inevitable, (Panko, 2005). Human beings commit errors in every walk of life. It is the very internal nature of human beings and is very difficult to change. But, what can be changed is the external nature. The idea is to modify the external factors to cope with the erring nature of humans, and thereby, improve the accuracy of spreadsheets. 'External factors' mean strategies or approaches incorporated by the organisations to maintain the quality of spreadsheets. To begin with, we mention some evidences of spreadsheet errors and the steps taken by some researchers to enhance the quality of spreadsheets.

Spreadsheet models are very widely used and are very likely to contain errors (Panko and Halverson, 2001). Following are some recent evidences of the occurrence of spreadsheet errors.

A simple spreadsheet error (cut-and-paste) cost a firm a whopping US$24m. The mistake led to TransAlta, a big Canadian power generator, buying more

US power transmission hedging contracts at higher prices than it should have. (Cullen, 2003)

A US government audit says the Columbia Housing Authority has to pay $216,352 to cover expenses incurred as it gave some Section 8 tenants too much room and landlords excess rent. Phil Steinhaus, the housing authority's CEO, asked that the fees for over-housing be waived but agreed to pay $118,387, the amount that resulted from a spreadsheet data-entry error that overpaid landlords. (Miller, 2006)

A chaotic situation in the posting of minimum bid prices for the first phase of North Port's abandoned lot auction led to confusion as the cost of some lots seemingly tripled overnight. In a rush to make the prices available to public before Christmas, the appraiser hired by the county put the auction lot number, the property ID number and the minimum bid amount onto a spreadsheet in sequential order but, inadvertently, did not sort the value column. (Venice Gondolier Sun, 2006)

Eastman Kodak Co. added $9m to its big third-quarter loss, to correct its several accounting errors. The adjustments reflect restructuring and severance costs linked to its ongoing effort to turn itself into a digital photography business. A Kodak spokesman said an $11m severance error was traced to a faulty spreadsheet and there were too many zeros added to the employee's accrued severance. But it was an accrual. There was never a payment. (Jelter, 2005)

A miscalculation in a spreadsheet almost cost Chi Omega sorority first place in the Homecoming competition. Katie Gonsoulin, Homecoming Committee chairperson, said the error occurred when the formula used to calculate scores from Homecoming Week events left two scores out of the tabulation. The resulting scores announced at the Homecoming game were incorrect. (Beagle, 2004)

Westpac had to halt trading on its shares and deliver its annual profit briefing a day early, after it accidentally emailed its results to research analysts. Details of the $2.818bn record annual profit result, which were due to be announced, were overshadowed by concerns of some information being leaked into market. The new figures were embedded in a template of last year's results and were accessible with minor manipulation of the spreadsheet. Chief financial officer, Philip Chronican, said it was not just one error, but a compounding of 2 or 3 errors. (Knight, 2005)

**1.3 Approaches by Other Researchers**
Rajalingham et al (2000) proposed an approach, the significant feature of which is that it adopts concepts from software engineering and employs important principles and techniques such as a unique definition of spreadsheet model elements (chiefly labels, data values and formulae), hierarchical representation of a formula in tree form, and separation of data (user-entered data values) and operations (formulae that operate on them).

Berge et al (2005) worked on a project to help end-users to locate and prevent, principally, mistyping and other human errors. Their implementation gives an option to visualize dependencies (represented by arrows) between cells in the spreadsheet to help the user see any inconsistencies in references between cells. Also, they implemented a way to assign a type to a cell which warns the user when a faulty type is entered. Further, they have a tool which visualizes the types and gives a better overview of the types in the spreadsheet. UML diagrams (Use-cases,

Class diagrams and Interaction diagrams) were used in the requirements planning and design phases of this project.

Aiming to facilitate analysis and comprehension of the different types of spreadsheet errors and to clearly understand the characteristics of an error as well as the nature of its occurrence, Rajalingham et al (2000) came up with a classification or taxonomy of errors. This is an outcome of a thorough investigation of the widespread problem of spreadsheet errors and an analysis of specific types of these errors. It also enables users to gain a better understanding of the different types of errors that can occur in their spreadsheet models. Appropriate tools, techniques and methods can subsequently be developed to prevent their occurrence in the first place or enhance the chances of detecting these errors after they have occurred. In addition to that, when a new specific type of error is identified, it can be placed in the appropriate category within the taxonomy. In the process of classifying the error, spreadsheet developers and end-users are bound to gain a much deeper understanding of the error. This is because they are forced to examine and compare its characteristics with those of other spreadsheet errors.

Another important strategy is 'code inspection'. Panko (1999 cited Panko 2005) found that team code inspection allowed undergraduate MIS majors to find 83% of all seeded errors in a spreadsheet, although the group did not find errors not previously found by the members of the team, who had inspected it alone before the group code inspection. Panko's study was centred on **'tetrads'** to detect errors seeded in spreadsheets already designed.

**1.4 Our approach to group work**
Contrastingly, our study as discussed below, is centred on working in 'pairs' to cross-check the overall work done individually. Our study also addresses several other aspects of spreadsheets with regard to design, implementation and testing: namely modelling, determining the appropriate formula to solve the problem, entering data into the cells and presenting the data. A novel aspect of our study is that **'dyads'** cross-checking their work could find errors unidentified when they worked on their own. Usefully, employees' perceptions on group work and on working in pairs to cross-check their work were also reported.

This study was based on an assumption from the evidences of spreadsheet errors that some errors might have been committed either in a hurry or due to lack of time to cross-check with others. Also, some errors could have probably been avoided if they had taken time and/or cross-checked with others. The following experiments were conducted:

1. Assessing the usefulness of cross-checking to improve spreadsheet accuracy.
2. Evaluating the benefits of group work and comparing it with the cross-check approach.
3. Examining the effects of time constraints on spreadsheet accuracy.

Surveys of spreadsheet developers (Panko, 2005) indicate that spreadsheet creation, in contrast, is informal, and few organizations have comprehensive policies for spreadsheet development. Further, as we have seen, there are diverse approaches like legal policies, software engineering and development techniques, group work and other strategies proposed by various researchers. However, the seriousness of spreadsheet errors justifies the necessity of varied approaches to enhance the spreadsheet quality. As with any true engineering discipline, spreadsheet engineering looks set to require numerous and distinct strategies to encompass such a troubling issue.

## 2. KNOAH SOLUTIONS

Knoah Solutions is a leading offshore outsourcing company with facilities in Hyderabad, India, providing multi-channel customer and technical support for technology products and services, thereby enabling US call centre quality at competitive offshore prices. Knoah's commitment to quality is demonstrated in their ISO 9001:2000 certification, (Knoah Solutions Pvt. Ltd., 2006). The basic qualification for an employee in Knoah is a bachelor's degree plus computer skills. Since these agents use MS Excel they were suitable candidates for the above experiments, which were conducted via a Team Leader at Knoah.

## 3. EXPERIMENT ONE (CROSS – CHECK APPROACH)

### 3.1 Aim
The aim was to determine if cross–checking of spreadsheets makes any difference to the accuracy, e.g. enhances accuracy. Employee's perceptions on this cross-checking approach were also sought.

### 3.2 Experiment Design
The experiment consists of two phases, which are described below.

**First Phase (Working Individually)**
The idea was to take a sample of volunteers and give them a task to complete in Excel. The task to be assigned (in all the experiments) could be a combination of any two or all of the following sub-tasks**:** entering a considerable amount of data (already supplied), performing certain operations (including constructing formulae) on the data, presenting the data entered and that generated by the formula graphically. Once, they finish the task given to them in a fair amount of time, they were to save the files and send them for evaluation.

**Second Phase (Cross-Checking in Dyads)**
Before proceeding to the second phase of the experiment, the spreadsheets received at the end of the first phase were checked for accuracy. Then, individual participants were paired up with respect to their validity. Possible pairing were correct with correct, correct with incorrect and incorrect with incorrect. The confidentiality of the validity of the solutions was maintained when the participants were paired up to compare and check their work for errors. After working in pairs, the participants produced final common solutions. Due to the pairing up process, the number of final solutions was exactly half the number of solutions received during the first phase. Lastly, little questionnaire was sent to the participants to seek their views and comments on the cross-check process.

### 3.3 Pilot Testing
Pilot testing is vital before conducting an experiment in order to avoid inappropriate results. Tests with two similar tasks were conducted on a sample of 6 known subjects with the objective of determining what is a reasonable amount of time to finish the task and then to cross-check the solution with another person.

During the pilot tests, the validity of the solutions in the first phase was intentionally kept secret when the participants were paired up to compare and check their work for errors. It is interesting to mention that in a pilot test, two of the participants who were right in the beginning ended up with a wrong solution after cross-checking. This was because of their under-confidence about applying the appropriate formula - they became confused by each other.

### 3.4 Conducting the Experiment
This experiment was done on a sample of 18 agents at Knoah Solutions. The same task used during the pilot tests was assigned to the agents and they were given 30 minutes to finish the same.

The task contained payroll information for Cardiff Supermarket Ltd. (a fictitious name) for the year, 2004-2005. The names of the staff members along with their designation or department to which they belong, their basic wage and overtime wage are listed. The task was to calculate the average wage per person in each department (or designation) and also to represent the department/designation and the respective average wage graphically. This task was adapted from a similar task involved in a study on 'Misconception of the AVERAGE function'. (Rajalingham et al, 2000).

In about 30-35 minutes after assigning the task, excel solutions were sent by all the agents using their corporate emails ids. These solutions were checked against the correct solution. Only 7 out of 18 came up with the correct solution (average pay). Among the 11 incorrect solutions 16 errors were identified.

After determining the validity of the solutions, it was decided to group the 18 agents into 9 pairs. Among them, 1 pair has to be formed by agents who were correct in the first phase, 3 pairs have to be formed by agents who were incorrect, and the rest (5 pairs) being a combination of both of them. The cross-check questionnaire was also sent along with the list of pairs of names who will cross check their work. And, in 18-20 minutes, 9 final common solutions and 18 answered questionnaires were emailed by the respondents. The Excel sheets thus received in the second phase were checked for errors. Only one solution was wrong. The only mistake in it was the usage of incorrect formula.

### 3.5 Results
**Accuracy Statistics**
In the first phase of the experiment, when the 18 agents were working on their own, only 7 finished the task correctly and the rest were incorrect. This means, the percentage of accuracy is 38.88%. In the second phase, when the 18 agents were grouped into 9 pairs and asked to review their work together, 8 out of the 9 pairs came up with correct solution. This, in effect, means 16 out of 18 agents were correct. That is, 9 out of 11 agents rectified their mistakes. So, the final percentage of accuracy is 88.88%.The increase in accuracy in the second phase over the first phase is 50% and the percentage increase in the accuracy is 128.60 %

**Employee's Perception on 'Cross-Check' Approach**
It appears from the responses in the questionnaires that most employees liked this idea of 'cross-checking'. All 7 employees who were right in both the phases expressed that this process helped in finding the errors and reassured them of the accuracy of their work. Among the remaining 11 agents who were wrong in the first phase, 7 stated that cross-checking is a helpful strategy. The rest, 4, were unsure about the benefits of this approach and it is understood from their responses that they chose 'Not Sure', as they were not confident about the validity of their final common solutions. So, overall, 77.77% of the participants found the 'cross-checking' idea beneficial, while the rest were unsure.

## 4. EXPERIMENT TWO (GROUP WORK)

### 4.1 Aim
The aim was to determine: if working in groups is as effective as working separately and then cross-checking in groups, in terms of increasing the spreadsheet accuracy. A further aim was to examine and compare the accuracies when individuals worked separately, in dyads and in triads. Again, the experiment sought employees' feedback on group work.

### 4.2 Experiment Design
This experiment involved *n* individuals (working separately), *n* dyads and *n* triads, all working on the same spreadsheet task simultaneously. Once the assigned time elapsed, the participants had to send in the spreadsheets through email. Individuals who worked in dyads or triads had to come up with only one solution for their respective dyad or triad. Again, the participants' response to working in dyads or triads was sought by a 'Yes' or 'No'. All the solutions received were evaluated and the percentage accuracy for the three groups was calculated for comparison.

### 4.3 Pilot Testing
This experiment was pilot tested on a sample of 14 Part-Time MBA students at the University of Wales Institute, Cardiff. (UWIC) They worked together in 7 pairs on the same task as that used in the first experiment. The time allotted for the task was 30 minutes. Both of the students in one of the dyads had little awareness of spreadsheet usage and so never finished it. So, only 6 common solutions were received, of which, 3 were wrong and 3 were right.

It was observed that the individuals working in pairs were sharing parts of the tasks between each other. That is, while one was reading out the values, the other was entering, while one was counting and adding up the numbers, the other was just typing in those calculated values dictated by the other and while one of them worked out the formula, the other implemented it. So, in effect, only one of the two students in the dyads seemed to be working. That implies if one of them is wrong the dyad is wrong. And very little or no effort was observed to be put by them to ensure if they were right.

### 4.4 Conducting the Experiment
This experiment was done with the support of a Team Leader at Knoah on a sample of 36 agents, of whom, 6 worked individually, 12 worked in pairs and 18 in groups of three. That is, there were 6 individuals, 6 dyads and 6 triads. The task used in this experiment is same as the one used in the first experiment. The time allotted was 30 minutes.

### 4.5 Results
**Accuracy Statistics**
Out of 6 individuals who worked separately, 3 were right and 3 were wrong. Among the 6 dyads, 3 were right while the rest were wrong. Of the 6 triads, 5 came up with the right common solutions, but only one triad sent a wrong solution. The percentage of accuracy for the group of individuals working separately was 50%.

This was the same as the accuracy for the group of dyads. But the percentage accuracy for the group of triads was 83.33%.

**Employees' Opinions on Group Work**
Among 12 agents who worked in pairs, 3 disliked it, while the rest liked working in dyads. Out of 18 who worked in groups of three, 3 agents disliked it and the rest liked working in triads. So, out of 30, (who worked either in dyads or triads) 24 liked group work and the remaining disliked it. The percentage of agents who liked working in dyads was 75% and for triads, it was 83.33%. The overall percentage of agents who liked group work was 80%.

**Observations during the Experiment**
There was a similar sort of behaviour of the agents (i.e. sharing work), as outlined in the pilot testing, observed by the Team Leader who conducted this experiment. Another interesting observation made was that there was more participation among the members who worked in triads than those who worked in dyads. This observation was further strengthened by the accuracy statistics mentioned above.

## 5. EXPERIMENT THREE (TIME CONSTRAINTS)

### 5.1 Aim
The aim of this experiment was to determine if time constraints imposed on completing spreadsheet tasks have any impact on the accuracy. Further, the aim was to examine the accuracy with decreasing time allowed, using various time limits.

### 5.2 Experiment Design
This experiment consists of 5 phases, each of which needed a day to be carried out. Five different spreadsheet tasks, with equal complexity and which took the same amount of time to complete, were used. The sample in all the phases was necessarily the same. The time duration assigned was progressively and evenly decreased. Once the assigned time elapsed in each phase, the participants had to submit their Excel sheets by email, no matter whether they were complete or not. All the solutions received were evaluated and the percentage accuracy in each phase was analysed.

### 5.3 Pilot Testing
The pilot tests were conducted using five tasks on a sample of eight known subjects. The objectives of the pilot experiments were: to determine suitable time limits to finish the tasks, to confirm the accuracy/validity, to ensure similar complexity in each task, and finally, to make sure the tasks consumed equal times. However, the objective was not to examine the accuracy with varying time limits.

### 5.4 Conducting the Experiment
This experiment was also conducted at Knoah Solutions on a sample of 19 agents. The phase-wise description of the experiment is given below.

**Day 1: Phase 1**
The time duration assigned to complete the task in this phase was 24 minutes. The task was based on a person's shopping (of 4 different fruits) for himself and his friends for Easter. The number of each different fruit he bought needed to be calculated from the information provided in the task. The question was to calculate the total number of fruits he can distribute to each of his six friends. The spreadsheets were submitted by the above 19 agents promptly, 24 minutes after assigning the task. These were checked for errors. Evaluation involved checking the final answer and the step-by-step explanation in arriving at this final value. Only 2 agents came up with incorrect solutions. The correct solutions for all the tasks were already worked out during the pilot testing, the printouts of which were taken so that it would be easy for evaluation. Evaluation in each phase involved comparing the values in the excel sheets (responses) against those in the printouts.

**Day 2: Phase 2**
The time duration allotted to finish the task in this phase was 20 minutes. This task was based on calculating the simple Interest, compound interest and the difference between them for a given list of customers, principle amounts, loan periods and interest rates. As requested, the participants submitted their solutions 20 minutes after receiving the task. The primary focus of evaluation in all the phases was on checking if the correct and relevant formula was used to calculate the required value. The difference in the compound and the simple interests were checked for all the 30 customers in each spreadsheet. While 5 of them were wrong, 14 of them were right.

**Day 3: Phase 3**
Third day, another task was sent to the agents, the time allotted was 16 minutes. The task used in this phase was to determine the heat energy by a combination of eight different calculations on (five) values recorded during various cases in a thermal power station. The agents acted accordingly and emailed their spreadsheets after the allotted time elapsed. Evaluation of their work involved checking the amount of heat energy for all the twenty cases given in the task. It was found that 10 agents were right and 9 were wrong.

**Day 4: Phase 4**
Next day, a different equally complex task was sent to the agents. Further, the task assigned in this phase was very similar to the one assigned in the previous phase. This also was centred on determining some scientific value by a combination of a variety of calculations on values recorded in various cases in an engineering plant. The time duration to complete the task was 12 minutes. The solutions received were then checked for accuracy. This involved checking the final scientific value for all the given cases, against the values already worked out. It was determined that among the 19 solutions, 9 were right and 10 were wrong.

**Day 5: Phase 5**
Last day, the task duration was 8 minutes. This task is same as the one used in the first experiment, except that no graph is required here. The number of incorrect solutions was 14 and that of correct solutions was 5 in this phase.

**5.5 Results**
**Accuracy Statistics**
The overall accuracy for the 19 agents in the 5 phases is represented in the Table 1.

| Time Duration | 24min | 20min | 16min | 12min | 8min |
|---|---|---|---|---|---|
| Correct Solutions | 17 | 14 | 10 | 9 | 5 |
| Incorrect Solutions | 2 | 5 | 9 | 10 | 14 |
| % of Accuracy | 89.47 | 73.68 | 52.63 | 47.37 | 26.32 |

Table 1

The percentage accuracy in all phases is graphically shown in the Figure 1.

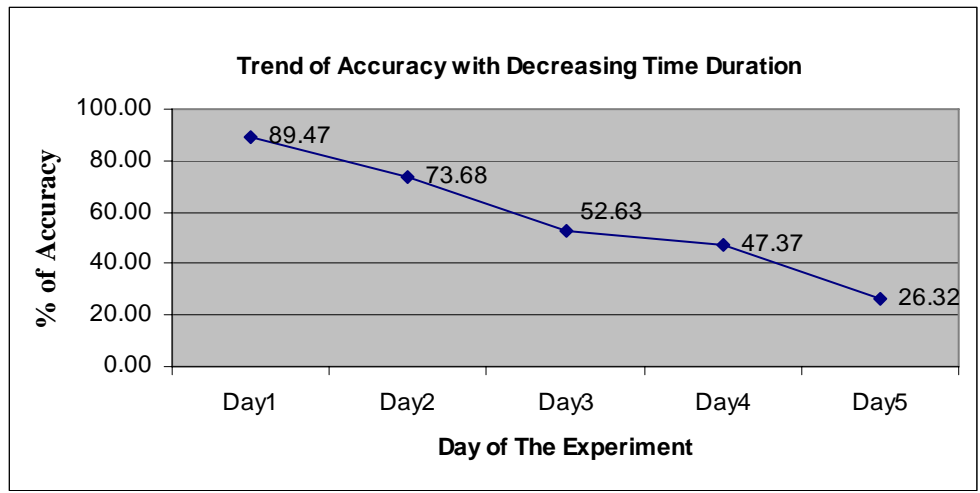

Figure 1

There is not much decrease in the accuracy from day 3 to day 4 because, as mentioned earlier, the tasks used on these days were quite similar. This was done purposely to identify if a 'learning process' has any influence on the accuracy.

## 6. CONCLUSION

The results from the first experiment make it clear that cross-checking of spreadsheets detects errors unidentified when users or developers work on their own. The accuracy in the second phase of the experiment is more than double that in first phase. And hence, the extent to which this approach improves the accuracy is undoubtedly, significant. Another important point to be noted is that more than three-fourth of the participants found this idea beneficial in ensuring accuracy and lessening the number of errors in spreadsheets.

In the second experiment, the lack of participation observed in the dyads could be due to an intentional or unintentional lack of interest and concentration, dependence on the partner. But this was not the case when they first worked individually and later cross-checked in dyads, as in the first experiment. Comparison of the accuracy statistics for both the experiments also suggests that working individually and then cross-checking in groups is a better approach than directly working in groups. The accuracy was same for the group of agents who worked individually and the group of dyads but was higher for the group of triads. Most of the participants preferred group work.

We conclude: considering the potential risks that the spreadsheet errors pose, it is worthwhile to assign multiple users to work separately on the same spreadsheet task and later cross-check with each other to assure accuracy. Effectively, this idea is a combination of 'individual work' and 'group work' hence claiming the advantages of both strategies. Overall, this strategy is justifiably suggested for crucial spreadsheets essential for business-critical decisions.

It appears from the results of the third experiment that time has a significant impact on the quality of spreadsheets. As the assigned time limit decreases, the accuracy drops proportionally. The time constraint rules the minds of the employees and builds pressure on them. Due to this, they cannot cope with any aspects in the task that are confusing and ultimately, make mistakes. Also, they cannot make time to review their spreadsheets. We conclude: while most organisations require their employees to get more work done in less time in order to cut costs, these restrictions would only result in poor quality of spreadsheets.

**6.1 Limitations to the Experiment**
The sample sizes used in these experiments are 18, 36 and 19 respectively. Perhaps, larger samples could have strengthened the conclusions. The results may also have been different, had the experiment been conducted on very highly skilled spreadsheet professionals. Also, more complex tasks might have yielded different results. As mentioned earlier, these experiments were carried out by a Team Leader at Knoah. Further observation of the behaviour of the agents could have been made during the experiments.

**6.2 Further Research**
Further research needs be done on the above limitations. This study could be extended to examine the accuracy by varying two or all the three of**:** time restrictions, complexity of the task and the number of users cross-checking their work in groups (triads, tetrads and pentads) once they finish working separately. It could also be a fetching idea to extend the research to include factors like experience and overconfidence.